\documentstyle[aaspp4,psfig,flushrt]{article} 
\begin{document}

\title{ON THE CORRELATION OF TORQUE AND LUMINOSITY IN GX~1+4}

\author{Deepto~Chakrabarty,\altaffilmark{1,2}
  Lars~Bildsten,\altaffilmark{3} Mark~H.~Finger,\altaffilmark{4,5}
  John~M.~Grunsfeld,\altaffilmark{1,6} Danny~T.~Koh,\altaffilmark{1}
  Robert~W.~Nelson,\altaffilmark{1} Thomas~A.~Prince,\altaffilmark{1} 
  Brian~A.~Vaughan,\altaffilmark{1} and Robert~B.~Wilson\altaffilmark{4}}  
\affil{\footnotesize deepto@space.mit.edu, bildsten@fire.berkeley.edu,
  finger@gibson.msfc.nasa.gov, jgrunsfe@ems.jsc.nasa.gov, 
  koh@srl.caltech.edu, nelson@tapir.caltech.edu,
  prince@caltech.edu, brian@srl.caltech.edu, wilson@gibson.msfc.nasa.gov}

\altaffiltext{1}{Space Radiation Laboratory, California Institute of
  Technology, Pasadena, CA 91125.}
\altaffiltext{2}{Current address: Center for Space Research,
  Massachusetts Institute of Technology, Cambridge, MA 02139.}
\altaffiltext{3}{Department of Physics and Department of Astronomy,
  University of California, Berkeley, CA 94720.}  
\altaffiltext{4}{Space Science Laboratory, NASA/Marshall Space Flight
  Center, Huntsville, AL 35812.}
\altaffiltext{5}{{\em Compton Observatory} Science Support Center,
  Goddard Space Flight Center/Universities Space Research Association.} 
\altaffiltext{6}{Current address: NASA/Johnson Space Center, Code CB,
  Houston, TX 77058.}

\smallskip
\centerline{\small To appear in {\sc The Astrophysical Journal Letters} (1997)}

\begin{abstract}
Over five years of daily hard X-ray ($>$20 keV) monitoring of the 2
min accretion-powered pulsar GX 1+4 with the {\em Compton Gamma Ray
Observatory}/BATSE large-area detectors has found nearly continuous
rapid spin-down, interrupted by a bright 200-d spin-up episode. During
spin-down, the torque becomes more negative as the luminosity
increases (assuming that the 20--60 keV pulsed flux traces bolometric
luminosity), the opposite of what is predicted by standard accretion
torque theory. No changes in the shape of the 20--100 keV pulsed
energy spectrum were detected, so that a very drastic change in the
spectrum below 20 keV or the pulsed fraction would be required to make
the 20--60 keV pulsed flux a poor luminosity tracer.  These are the
first observations which flatly contradict standard magnetic disk
accretion theory, and they may have important implications for
understanding the spin evolution of X-ray binaries, cataclysmic
variables, and protostars.  We briefly discuss the possibility that
GX~1+4 may be accreting from a retrograde disk during spin-down, as 
previously suggested.
\end{abstract}


\section{INTRODUCTION} 

The torque exerted on a magnetic star by an accretion disk is of great
interest in astrophysics, with relevance to binary evolution, star
formation, neutron star structure, and the origin of millisecond radio
pulsars.  Accretion-powered pulsars are the ideal laboratory for the
study of accretion torques, since the bolometric X-ray intensity is a
tracer of the mass accretion rate, and the X-ray pulsations and the
neutron star's small moment of inertia permit torque measurements on
short ($\sim$days) time scales.  For accretion from a prograde disk,
the material torque will generally act to spin up the star until it
reaches its equilibrium spin period, where steady accretion is halted
by a centrifugal barrier (Pringle \& Rees 1972; Lamb, Pethick, \&
Pines 1973; Illarionov \& Sunyaev 1975). This shutoff occurs when the
magnetospheric radius $r_{\rm m}$ (where the Keplerian kinetic stress
is equal to the magnetic stress) is comparable to the corotation
radius $r_{\rm co}$ (where the magnetic field lines move at the local
Kepler velocity).

Early observations of disk-fed X-ray pulsars found that the simple
model for steady spin-up sketched above is sometimes inadequate: some
X-ray pulsars spin up at rates much smaller than predicted, or even
spin down for extended intervals while continuing to accrete
matter. This led to suggestions that additional magnetic spin-down
torques must be present, capable of reducing or even dominating the
material spin-up torque even while the star continues to
accrete. These models invoke magnetic coupling of the accretion disk
and the magnetosphere (Ghosh \& Lamb 1979a, 1979b; Wang 1987, 1995) or
loss of angular momentum through the expulsion of a
centrifugally-driven magnetohydrodynamic wind (Anzer \& B\"orner 1980;
Arons et al. 1984; Lovelace, Romanova, \& Bisnovatyi-Kogan 1995). All
of these near-equilibrium disk-accretion scenarios predict that a
higher mass accretion rate $\dot M$ should yield a smaller
magnetospheric radius $r_{\rm m}$ and a larger spin-up torque. They
likewise predict that a reduced $\dot M$ increases $r_{\rm m}$ and
reduces the accretion torque until a net spin down occurs. For
sufficiently low $\dot M$, this can eventually lead to centrifugal
inhibition of accretion due to a ``propeller'' effect (cf. Illarionov
\& Sunyaev 1975).

Most of the sparse, intermittent observations of X-ray pulsars during
the 1970s and 1980s were generally consistent with the
near-equilibrium scenario (Nagase 1989 and references therein).
However, more recent long-term monitoring of a large sample of
accreting pulsars with the Burst and Transient Source Experiment
(BATSE) on the {\em Compton Gamma Ray Observatory} has found several
systems whose behavior is difficult to explain in this context
(Bildsten et al. 1997).

One of the most interesting tests of accretion torque theory has come
from observations of the 2-min X-ray pulsar GX~1+4. This long-period
($\gtrsim 200$~d) symbiotic binary also contains the M6 III giant V2116
Oph and an accretion disk (Davidsen, Malina, \& Bowyer 1977;
Chakrabarty \& Roche 1997).  Throughout the 1970s, GX~1+4 was
consistently bright ($\gtrsim 100$ mCrab in the 2--10 keV band; see
McClintock \& Leventhal 1989 and references therein) and spinning up
rapidly with a mean rate $\dot\nu \approx 6.0\times
10^{-12}$~Hz~s$^{-1}$ (Nagase et al. 1989 and references therein).
There were no observations of GX~1+4 between 1980 and 1983, but
several observations by {\em EXOSAT} in 1983 and 1984 failed to detect
it, indicating an X-ray flux decrease of at least two orders of
magnitude ($\lesssim 0.5$ mCrab in the 2--10 keV band; Hall \&
Davelaar 1983, Mukai 1988). 

The pulsar reappeared with a reversed (spin-down) accretion torque and
low luminosity in 1987 (3 mCrab in the 1--30 keV band; Makishima et
al. 1988).  During 1989--1991 it continued to spin down rapidly with a
mean rate $\dot\nu \approx -3.7\times 10^{-12}$~Hz~s$^{-1}$, similar
in magnitude to the previous spin-up rate.  A surface dipole magnetic
field of nearly $10^{14}$ G is required for this slow pulsar if its
torque reversal is a sign of being near its equilibrium spin
period. GX 1+4 has the hardest X-ray spectrum among the persistent
X-ray pulsars. Since some authors have suggested that the location of
the high-energy break observed in X-ray pulsar spectra may be related
to the magnetic field strength (e.g., Pravdo et al. 1978, Tanaka
1986), this might be evidence of a strong magnetic field in GX 1+4.
Alternatively, Makishima et al. (1988) and Dotani et al. (1989)
suggested that the spin-down may be due to accretion from a retrograde
disk formed from the stellar wind of the red giant companion.

Long-term BATSE monitoring of GX 1+4 since 1991 has detected 
surprising changes in the apparent torque-luminosity relation which
are not readily understood in terms of a near-equilibrium accretion
torque model. In this {\em Letter}, we present the torque and
flux data for GX 1+4.  A preliminary account of this work was
presented by Chakrabarty (1996).

\section{OBSERVATIONS AND ANALYSIS}

BATSE is a nearly continuous all-sky monitor of 20~keV--1.8~MeV hard
X-ray/$\gamma$-ray flux (see Fishman et al. 1989 for a description).
Our standard pulsed source detection and timing analysis uses the
20--60 keV channel of the 1.024 s DISCLA data (see Chakrabarty et al.
1993; Chakrabarty 1996). The barycentric pulse frequency history of GX
1+4 from 1991 April 16 to 1997 January 17 (MJD\footnote{Modified
Julian Date = JD $-$ 2,400,000.5} 48362--50465) was determined by
dividing the data into five-day segments and searching the Fourier
power spectrum of each segment for the strongest signal in the pulse
period range 110~s$\la P_{\rm pulse}\la$~130~s.  Figure~1 shows the
long-term pulse frequency history of GX~1+4, including pre-BATSE
archival data.  The overall BATSE pulse frequency history shows a
significant quadratic trend towards smaller $\dot\nu$ magnitudes, with
$|\dot\nu/\ddot\nu|
\approx 10$~yr. There are also occasional oscillatory excursions about
this quadratic trend in the pulse frequency. These are too large to be
due to orbital Doppler shifts but might represent accretion torque
variations.

Mean pulsed energy spectra were measured by folding long segments of
the BATSE CONT data (16 energy channels at 2.048 s resolution) using a
pulse timing model derived from the pulse frequency history.  A
single-harmonic pulse model was employed to measure the pulsed count
rates in the resulting pulse profiles (see Chakrabarty et
al. 1995). The spectrum was measured during three different source
states: a relatively quiescent spin-down interval during 1993 January
23--April 23 (MJD 49010--49100), a portion of a bright spin-down
flare during 1993 September 9--21 (MJD 49239--49251), and the bright
spin-up interval during 1994 October 13--1995 March 22 (MJD
49638--49798). All three intervals were adequately fit by a thermal
bremsstrahlung model (including Gaunt factor) with $kT\approx 45$
keV. During the bright spin-up interval, pulsations were clearly
detected at photon energies as high as 160 keV.

We obtained a 20--60 keV pulsed flux history by folding five-day
intervals of the DISCLA data using the pulse periods determined from
Fourier analysis, measuring the phase-averaged pulsed count rate by
fitting the pulse profiles with a sinusoidal template (Chakrabarty et
al. 1995), and correcting the resulting pulsed count rates for the
instrumental response assuming a fixed source spectrum. Because GX 1+4
is faint relative to the background and the BATSE detectors are
uncollimated, this analysis technique is only sensitive to the pulsed
component of the 20--60 keV emission. Due to its faintness and source
confusion problems near the Galactic center, BATSE monitoring
measurements of GX~1+4 using Earth occultation techniques (Harmon et
al. 1992) are not available, so that its unpulsed component is
indistinguishable from the background. The top panel of Figure 2 shows
the hard X-ray pulsed flux history measured since 1991.  The detection
threshold of these 5-day integrations is approximately $1.5\times
10^{-10}$ erg cm$^{-2}$ s$^{-1} \approx 15$ mCrab (20--60 keV).  Some
of the points in the flux history are upper limits with this
approximate value.

To compute the torque history of GX 1+4, we calculated a running
3-point numerical derivative of the pulse frequency history.  For
those intervals prior to 1996 August where the pulsar was not
detected, the mean pulse frequency derivative was estimated using a
linear interpolation of the nearest pulse frequency measurements
adjacent to the undetected interval.  We did not attempt to
interpolate over the long undetected interval during 1996
August--December.  The resulting pulse frequency derivatives $\dot\nu$
(which are proportional to the net torque on the neutron star) are
shown in the bottom panel of Figure~2.

\section{TORQUE AND LUMINOSITY BEHAVIOR}

Our observations of GX 1+4 divide naturally into five states. During
MJD 48362--49613 (1991 April 16--1994 September 18), we observed continued
spin-down, with relatively steady, persistent 20--60 keV pulsed
emission of $\approx 2\times 10^{-10}$ erg cm$^{-2}$ s$^{-1}$
interrupted by intermittent bright flares of $\sim 20$~d duration. The
cross-correlation function (e.g., Bendat \& Piersol 1986) of the flux
and torque histories over this interval has a very strong negative
peak at zero lag with a correlation coefficient of $-0.85$.  This
indicates a strong anticorrelation (negative correlation) of the two
quantities, rather than the positive correlation predicted by standard
disk accretion torque theory.  This anticorrelation is clearly evident
in Figure 2 from the enhanced spin-down episodes accompanying most of
the intensity flares in this interval (MJD 48393, 48546, 48998,
49238). However, two of the weaker flares (MJD 49448, 49563) do not
show an obvious anticorrelation, and indeed seem to be accompanied by
{\em reduced} spin-down. 

We examined the torque-luminosity relation during this first spin-down
state by assuming $-\dot\nu\propto F_{\rm x}^\beta$ and fitting the
data to this model using a doubly-weighted regression (Feigelson \&
Babu 1992; Press et al. 1992). Excluding non-detections ($F_{\rm
x}\lesssim 1.5\times 10^{-10}$ erg cm$^{-2}$ s$^{-1}$) and points
where the torque was consistent with zero, the best fit power law
index is $\beta=0.48\pm 0.01$. The data and best-fit model are shown
in Figure 3. The scatter for points with $-\dot\nu < 4\times 10^{-12}$
Hz s$^{-1}$ is very large, with no clear correlation
present. The data from the weak flares on MJD 49448 and 49563 form the
extremely discrepant group of points in the lower right portion of
Figure 3. Excluding all of the points with $-\dot\nu < 4\times 10^{-12}$
Hz s$^{-1}$ does not significantly alter the best-fit $\beta$.

The second state began around MJD 49638 (1994 October 12), when GX~1+4
entered a prolonged ($\sim 200$~d) bright period. During this bright
state, the pulsar underwent a smooth torque reversal to spin-up. When
the pulsar began to fade, it also resumed rapid spin-down.  The
cross-correlation function of the flux and torque histories during MJD
49638--49813 (1994 October 12--1995 April 6) contains a strong
positive peak at zero lag, with a correlation coefficient of $+0.56$.
Because the spin-up episode spans only a factor of two in flux, it was
not possible to characterize the functional form of the
torque-luminosity relation.  The abrupt large spin-down torque
accompanying the low flux level around MJD 49825 (1995 April 18) is
suggestive of ``propeller'' spin-down due to the centrifugal
inhibition of accretion.

The third state began after the end of the bright spin-up episode.
During MJD 49848--50243 (1995 May 11--1996 June 9), the pulsed hard X-ray
emission varied erratically although the pulsar continued to spin down
at a relatively steady rate.  The wildly fluctuating flux history
suggests that the flux was varying on a time scale much shorter than
our 5-day measurements. The cross-correlation function of the torque
and flux histories during this interval contains a weak negative peak
at zero lag, with a correlation coefficient of $-0.87$.  The erratic flux
behavior was abruptly ended around MJD 50243 by a quiet interval
of about a month in duration, during which the pulsar was usually
below the BATSE detection threshold but evidently continued to spin
down.

GX 1+4 reappeared briefly with a bright flare peaking on MJD 50303
(1996 August 8), during which the pulsar made a transition to
spin-up. It then rapidly dropped below our detection threshold again
on MJD 50313 (1996 August 18), remaining undetected until MJD 50418
(1996 December 1).  The drop in flux was too rapid to allow us to
determine the torque state of the pulsar at the end of the flare.
This was the longest undetected interval for GX 1+4 since the start of
BATSE monitoring in 1991.  During this BATSE low state, a series of
2--60 keV observations with the {\em Rossi X-Ray Timing Explorer
(RXTE)} also failed to detect pulsations, establishing an upper limit
($\lesssim 0.2$ mCrab; Cui \& Chakrabarty 1996) comparable to the
1983/1984 {\em EXOSAT} low state.  Since MJD 50418, the pulsar has
been continuously detected by BATSE as of MJD 50465 (1997 January 17)
and is spinning down rapidly at a rate of $-7.5\times 10^{-12}$ Hz
s$^{-1}$.

\section{DISCUSSION}

According to standard accretion torque models, spin-down of a pulsar
accreting from a prograde disk requires that the neutron star is
rotating very near its equilibrium spin period, where $r_{\rm
m}\approx r_{\rm co}$. The primary evidence in favor of this
explanation is that, on average, the flux from GX 1+4 measured by
BATSE is higher during spin-up than during spin-down, as the standard
models predict.  The precise torque-luminosity relationship for the
pre-BATSE data is not known, because those observations did not have a
sufficient time baseline to measure the instantaneous torque
accurately.  Still, on average, the flux during the 1970s spin-up era
was significantly higher than during the 1980s spin-down era (e.g.,
McClintock \& Leventhal 1989), also supporting the equilibrium spin
explanation for GX 1+4. If this explanation is correct, then GX 1+4
must have the strongest known magnetic field of any neutron star.

However, the apparent anticorrelation of torque and luminosity
observed by BATSE during spin-down is inconsistent with the
near-equilibrium accretion torque models. Moreover, the flux during
the bright spin-down flare centered at MJD 49238 reached similar
intensities as the bright spin-up interval near MJD 49700 without
triggering a similar torque reversal, even though the duration of the
spin-down flares were much longer than the viscous time scale ($\sim$
hours) required for the inner disk to adjust to a higher $\dot M$.
Since BATSE only detects the pulsed hard X-ray emission from GX~1+4,
it is possible that our flux measurements are a poor tracer of
bolometric luminosity (which should be proportional to the the mass
accretion rate).  The observed changes in flux might actually be due
to changes in the spectral shape and/or pulsed fraction.  However,
since 20--100 keV pulsed spectra measured during three different
states show no significant difference in shape, any such spectral
changes must only occur below 20 keV.  Moreover, such changes would
have to be quite drastic in order to offset the factor of 7 increase
in BATSE flux measured during the MJD 49238 spin-down flare.  Multiple
observations with wide-bandpass missions like {\em RXTE} and the
European {\em Beppo-SAX} will eventually resolve this question.

Another puzzling observation is that the spin-up and spin-down torques
in GX 1+4 are quite similar in magnitude, and the transition between
the two states is relatively abrupt. Similar behavior has been
observed in several other accreting pulsars (4U 1626--67, Chakrabarty
et al. 1997; Cen X-3, Finger et al. 1994; and OAO 1657--415, Chakrabarty
et al. 1993), spanning a wide range in binary parameters and companion
types. The near-equilibrium accretion torque models do not explain why
a particular torque magnitude should be preferred in both spin-up and
spin-down.  They would require a sudden and finely-tuned transition in
$\dot M$. 

Based on the apparent torque-luminosity anticorrelation in GX 1+4 and
the similarity of spin-up/spin-down torque magnitudes in several
systems, Nelson et al. (1997) have revived an earlier
suggestion (Makishima et al. 1988, Dotani et al. 1989) that the
observed spin-down episodes may simply be due to the formation of a
retrograde accretion disk, with material rotating with the opposite
sense as the pulsar. As long as the magnetosphere is far inside the
corotation radius ($r_{\rm m} \ll r_{\rm co}$), a reversed disk should
produce a spin-down torque of similar magnitude as the spin-up torque
from a prograde disk. The torque magnitude in both cases would increase
with the accretion rate. For this explanation to be viable, the
switching of the accretion disk probably must occur on time scales
shorter than the time required to reach spin equilibrium, since
otherwise the spin-up and spin-down torque magnitudes need not be 
similar.

In the particular case of GX 1+4, the retrograde disk scenario for
spin-down would eliminate the need for an unusually strong magnetic
field, since the pulsar would necessarily be far from its equilibrium
spin period.  The chief question is how a retrograde disk could form and
remain stable over the long time scales of months or years over which
spin-down is observed. For systems in Roche lobe overflow, the
accretion stream carries high specific angular momentum with the same
sense as the orbit, so formation of a stable retrograde disk seems
implausible. However, GX 1+4 has several attributes which set it apart
from the other systems: a mass ratio near unity, a long ($\sim$years)
binary period, and a mass donor which may not be in Roche lobe
overflow and probably has a relatively dense, slow wind (Chakrabarty
\& Roche 1997). Under these conditions, formation of a retrograde
accretion disk may be possible.

\acknowledgements{We thank an anonymous referee for helpful comments.
This work was funded in part by NASA grants NAG 5-1458 and
NAGW-4517. D.C. was supported at Caltech by a NASA GSRP Graduate
Fellowship under grant NGT-51184. The NASA Compton Postdoctoral
Fellowship program supported D.C. (NAG 5-3109), L.B. (NAG 5-2666), and
R.W.N. (NAG 5-3119). L.B. was also supported by Caltech's Lee
A. DuBridge Fellowship, funded by the Weingart Foundation; and by the
Alfred P. Sloan Foundation.}

\vspace{1in}

\centerline{FIGURE CAPTIONS}

\noindent
FIG. 1.---The long-term pulse frequency history of GX 1+4. The
interval during which {\em EXOSAT} failed to detect the source is
indicated by the dotted lines.  The low state established by BATSE and
{\em RXTE} non-detections is also indicated. The BATSE measurements
comprise the solid line starting in 1991, and continuing after the
late-1996 low state.

\noindent
FIG. 2.---{\em Upper panel:} BATSE 20--60 keV pulsed flux history for
GX 1+4, averaged at 5-day intervals. The detection threshold is
$1.5\times 10^{-10}$ erg cm$^{-2}$ s$^{-1}$. Points whose error bars
intersect zero flux are upper limits. {\em Lower panel:} Pulse
frequency derivative ($\dot\nu$) history for GX~1+4. The gap during
MJD 50313--50418 corresponds to the BATSE and {\em RXTE}
non-detections.  The epochs of flares and other interesting events
discussed in the text are marked by the vertical dotted lines in both
panels.

\noindent
FIG. 3.---Torque-luminosity anti-correlation of GX 1+4 during the
1991--1995 (MJD 48362--49613) spin-down interval. Negative pulse
frequency derivative is plotted as a function of the 20--60 keV pulsed
flux.  Non-detections ($F_{\rm x}\lesssim 1.5\times 10^{-10}$ erg
cm$^{-2}$ s$^{-1}$) and points with torque consistent with zero are
omitted.  The dotted line indicates the best-fit power-law model for
the torque-luminosity relation, $-\dot\nu\propto F_{\rm x}^{0.48}$.
The data for smaller torques ($-\dot\nu< 4\times 10^{-12}$ Hz
s$^{-1}$) is very scattered and shows no clear correlation with flux.


\pagebreak
\pagestyle{empty}
\thispagestyle{empty}
\begin{figure}
\vspace{-0.3in}
\centerline{Figure 1}
\vspace{0.1in}
\centerline{\psfig{file=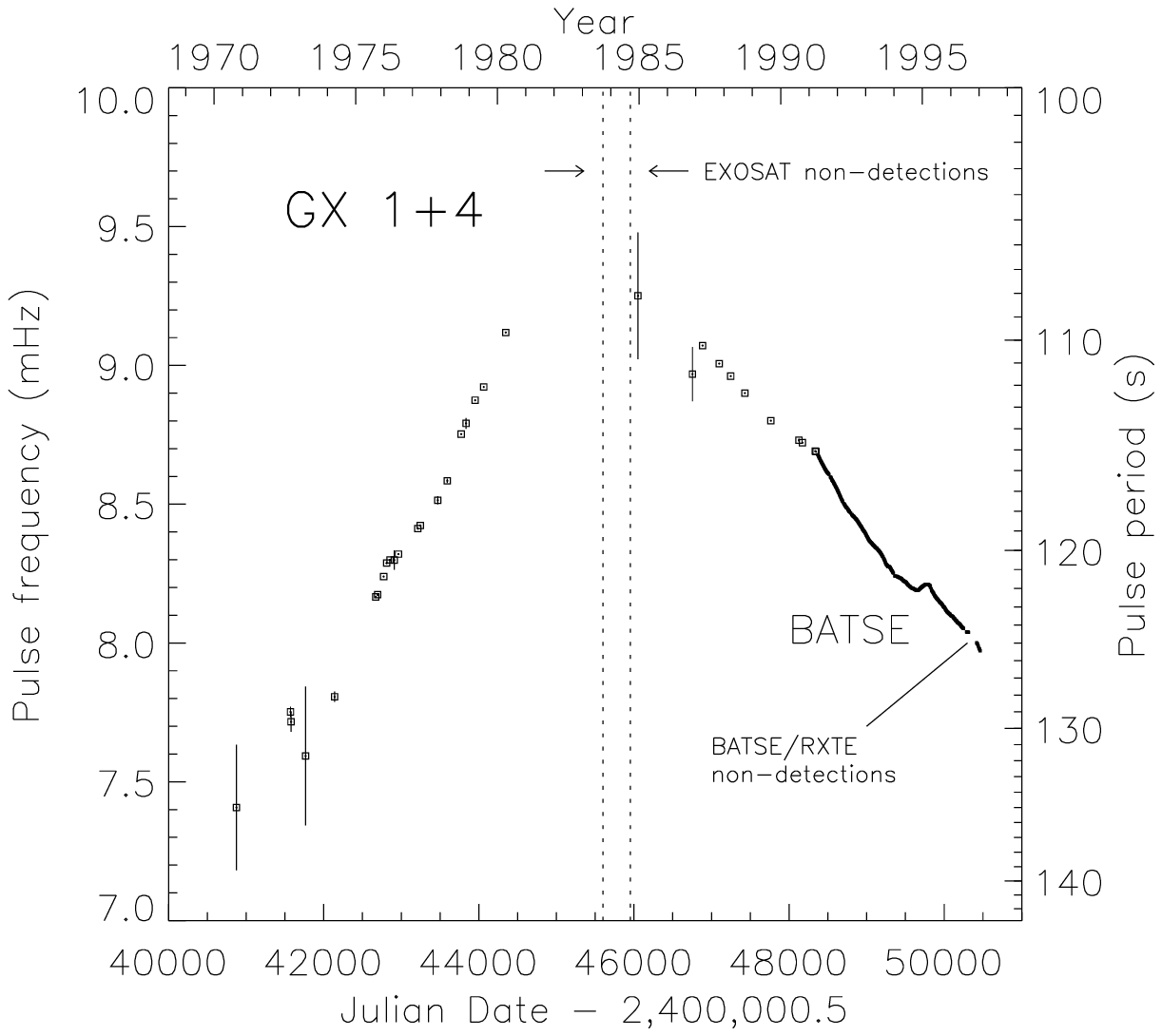}}
\end{figure}

\pagebreak
\pagestyle{empty}
\thispagestyle{empty}
\begin{figure}
\vspace{-0.4in}
\centerline{Figure 2}
\vspace{0.1in}
\centerline{\psfig{file=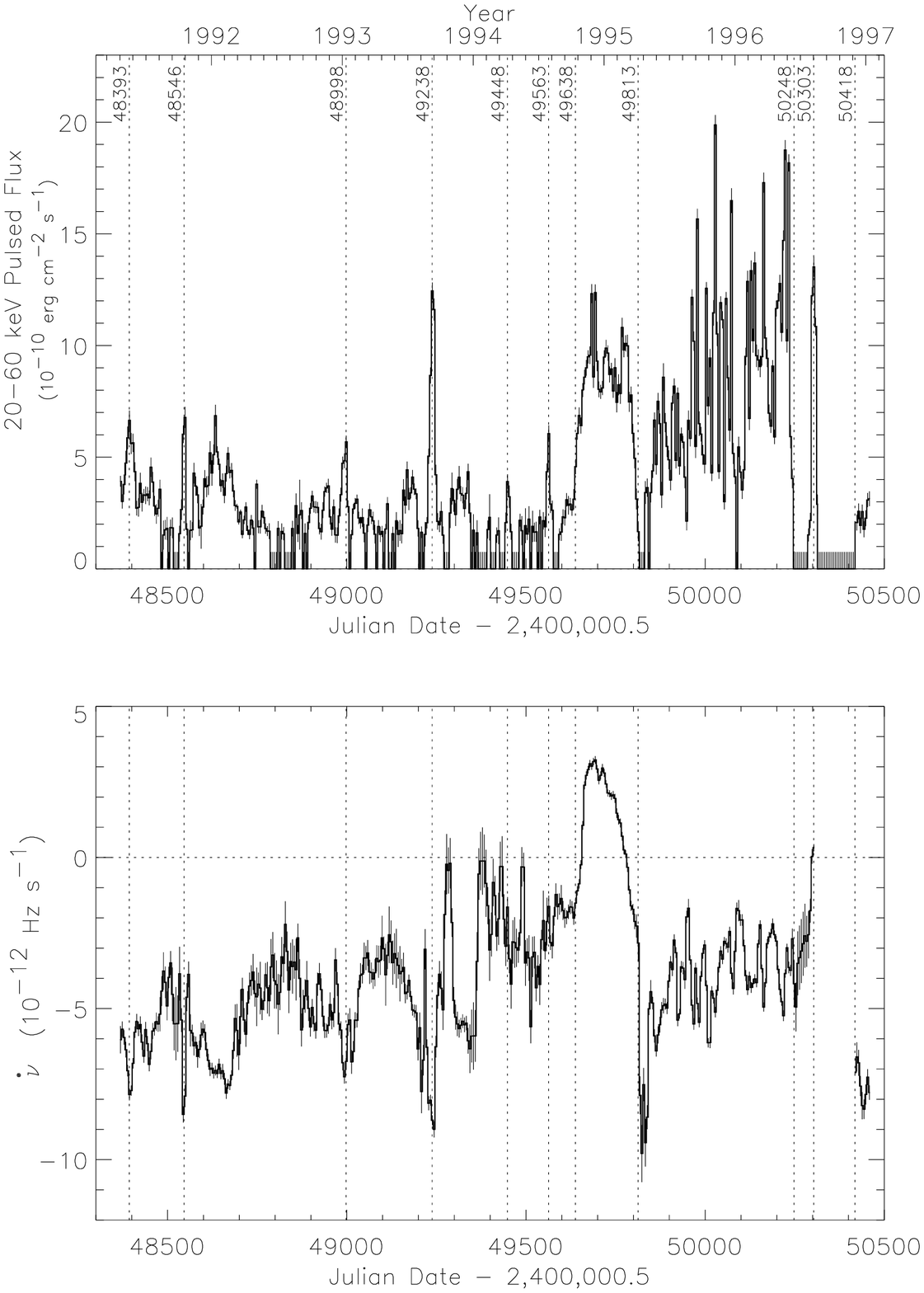}}
\end{figure}

\pagebreak
\pagestyle{empty}
\thispagestyle{empty}
\begin{figure}
\centerline{Figure 3}
\centerline{\psfig{file=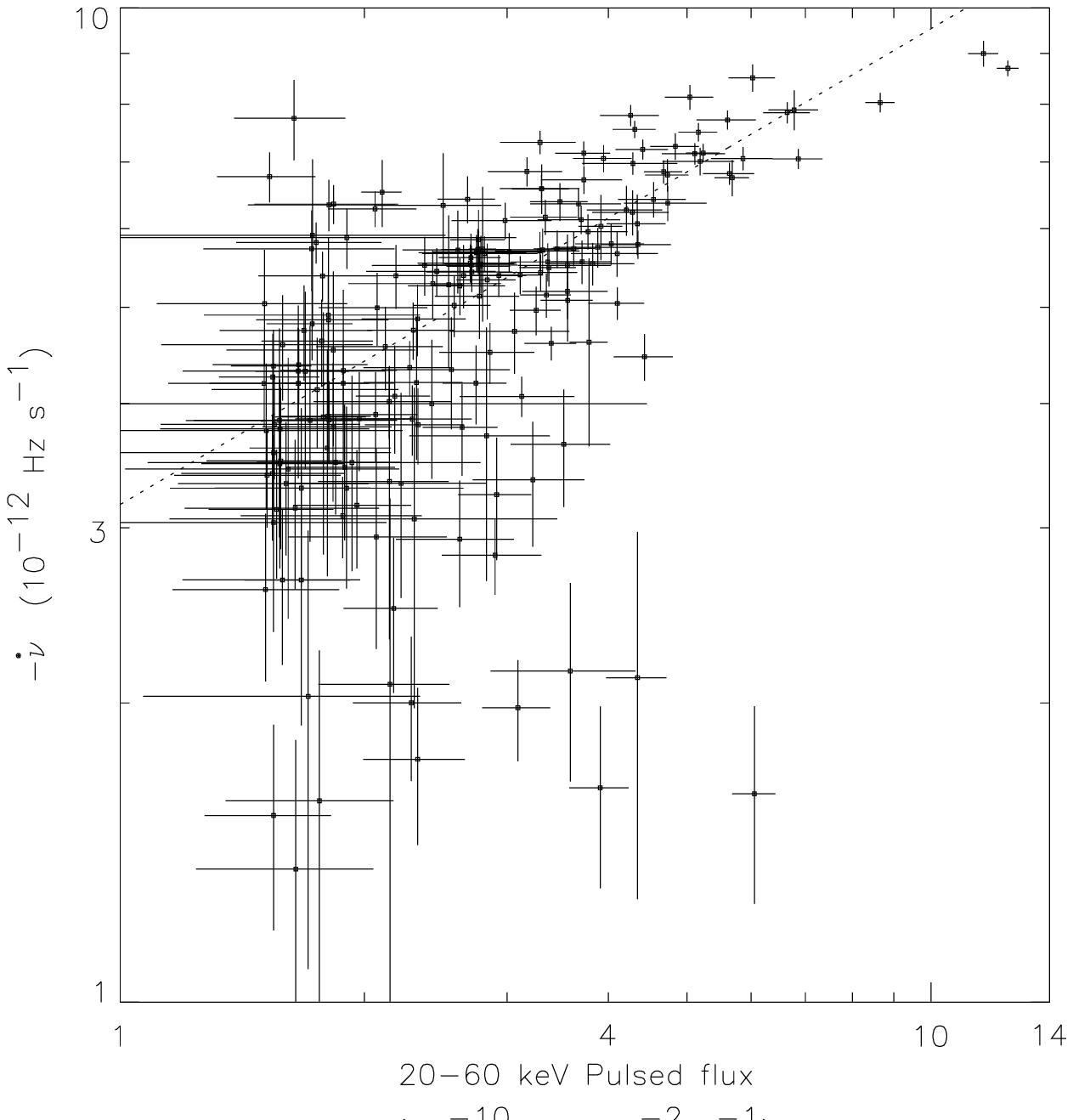}}
\end{figure}

\end{document}